# A Review Paper on Oculus Rift-A Virtual Reality Headset

Parth Rajesh Desai[1], Pooja Nikhil Desai[1], Komal Deepak Ajmera[1], Khushbu Mehta[1],

1. *U.G. student, Electronics and Telecommunication Department, DJSCOE, Vile-Parle (W), Mumbai – 400056, India.*

**Abstract: Oculus rift: Virtual reality (VR) is a burgeoning field that has the inherent potential of manipulating peoples mind with a superlative 3D experience. Oculus rift is one such application that assists in achieving the same. With the fleeting enhancements in VR it now seems very feasible to provide the user with experiences that were earlier thought to be merely a dream or a nightmare.**

## I. INTRODUCTION

Virtual Reality is a computer simulated environment that gives the user the experience of being present in that environment. It is a 3-Dimensional computer generated environment. VR provides the effects of a concrete existence without actually having a concrete existence. VR not only provides immersions of vision but also of sound and tactile feedback.

Basically, VR is a theory based on the human desire to escape the real world boundaries and this is done by embracing the cyber world. It is a new form of human machine interaction that is beyond keyboard, mouse or even touch screen for that matter. It is a means by which one can interact with full visual immersion. Immersion is based on two main components: depth of information and breadth of information. Depth of information includes resolution, quality and effectiveness of audio visuals etc. Breadth of information is the number of sensory present at a time. VR is implemented by using interactive devices like gloves headsets or helmets. Oculus Rift is a VR Ski-Masked Shaped Goggle device that works along with computers or mobiles. Other VR headsets have problem of motion sickness to the user post its usage. But here is a new technology which claims to have solved this problem of motion sickness and dizziness post the usage. This new technology is oculus rift.

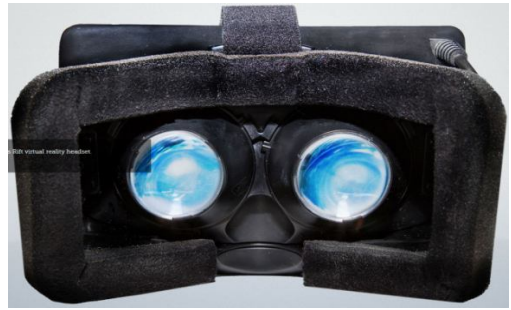

Fig. 1 The oculus rift headset

## II. DESCRIPTION

The Oculus Rift is a light weight headset that allows a user to step into the game and look in any direction. The Oculus Rift is currently present in developer kits versions. Two developer kit versions have been released namely Developer Kit version 1 (DK1) and developer kit version 2 (DK2). The DK2 is a more advanced version of the oculus rift as compared to DK1.

### A. Oculus Rift DK1 Specs

- Oculus Rift Price: $300 (INR 18000)
- Display Resolution: 640 x 800 (1.25:1) [1280 x 800 split between each eye]
- Display Technology: LED
- Display Size (diagonal): 4.08″
- Field of View (degrees): H: 90°
- Pixels Per Inch (PPI): 251
- Total Pixels (per eye): 512,000
- Weight (headset): 380g
- Stereoscopic 3D capable: Yes
- Audio: Bring Your Own Audio
- Inputs: DVI, USB (power)

- Head Tracking: Yes
- Positional Tracking: No

  *B.  Oculus Rift DK2 Specs*

- Oculus Rift DK2 Price: $350 (INR 21000)
- Display Resolution: 960 x 1080 (1.13:1) [1920x 1080 split between each eye]
- Display Technology: OLED
- Field of View (degrees): H: 90°
- Pixels Per Inch (PPI): 441
- Total Pixels (per eye): 1,036,800
- Weight (headset): 440g
- Stereoscopic 3D capable: Yes
- Audio: Bring Your Own Audio
- Inputs: HDMI 1.4b, USB, IR Camera Sync Jack
- Head Tracking: Yes
- Positional Tracking: Yes
- Refresh Rate: 75 Hz, 72 Hz, 60 Hz
- Persistence: 2 ms, 3 ms, full
- Viewing Optics: 100° Field of View (nominal)
- Cable: 10' (detachable)
- Input: HDMI 1.4b
- USB device: USB 2.0
- USB host: USB 2.0 (requires DC Power Adapter)
- Camera USB: USB 2.0
- Sensors: Gyroscope ,Accelerometer, Magnetometer
- Update Rate: 1000 Hz
- Sensor: Near Infrared CMOS Sensor
- Update Rate: 60 Hz

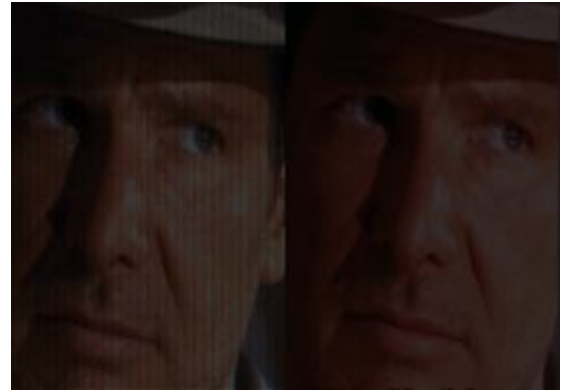

Fig. 2 Difference of quality of display between DK1 and DK2,  Left side image is of DK1 and right side image of DK2.

## III.    INSIDE VIEW

The inside view of the oculus rift goggles is as shown in the figure.

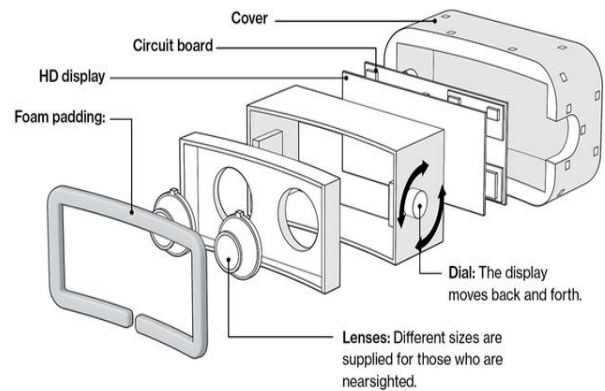

Fig. 3 The internal structure of oculus rift headset

The oculus rift kits come with three sets of lenses-A, B and C. lens pair A is to be used by people who have excellent long sighted eyesight as the rift is focused at infinity. The pairs B and C are to be used by people having problems with near sightedness, though cannot be used by all, especially people with major vision complications.

Also precise care should be taken while changing the lenses. A tiny dust particle if gets settled on the lens creates a dead pixel view in the VR. To worsen it further one can expect dust particles in both the lenses at different locations.

Furthermore glasses can also be worn along with the oculus rift goggles, provided that the glasses are not huge.

## IV.    WORKING

### A.   Head tracking

The oculus rift head tracking lets the user look around the virtual world just in the manner they would in the real world. The Oculus Rift head tracker constantly analyzes the player's head movement and uses it to control the view, instead of relying on a mouse or analogue stick to turn your view in the game. This makes for a completely natural way to observe the world, which is a major factor in immersion. The headset consists of a Oculus Tracker v2 board consisting of chips controlling the head tracking device. The chips are STM microelectronics 32F103C8 ARM Cortex-M3 Micro-controller with 72MHz CPU. Invensense .MPU-6000 six-axis motion tracking controller .Honeywell HMC5983 three axis digital compasses used in conjunction with the accelerometer to correct for gyroscope drift. The Oculus VR sensor, support sampling rates up to 1000 Hz, which minimizes the time between the player's head movement and the game engine receiving the sensor data to roughly 2 milliseconds. The increased sampling rates also reduce orientation error by providing a denser dataset to integrate over, making the player's real-world movements more in-sync with the game. The Oculus VR sensor includes a gyroscope, accelerometer, and magnetometer. When the data from these devices is fused, helps in determining the orientation of the player's head in the real world and synchronize the player's virtual perspective in real-time. The Rift's orientation is reported as a set of rotations in a right-handed coordinate system as given in figure(3) the gyroscope, which reports the rate of rotation (angular velocity) around X, Y and Z axes in radians/second, provides the most valuable data for head orientation tracking. By constantly accumulating angular velocity samples over time, the Oculus SDK(system development kit) can determine the direction of the Rift relative to where it began. Although the gyroscope provides orientation relative to the starting point, it creates two difficulties  it cannot provide the original orientation of the headset and it's subject to a small amount of drift over time (imagine re-orienting your head back to perfect center but in-game you're now looking slightly left or right).These issues affect the VR game with a fixed reference point (a game with a cockpit, where the players head's orientation does not affect the position of car/plane being piloted) .The accelerometer is leveraged to estimate the "down" vector and the magnetometer measures the strength and direction of the magnetic field. Combined, these allow for correction of drift in all three axes.

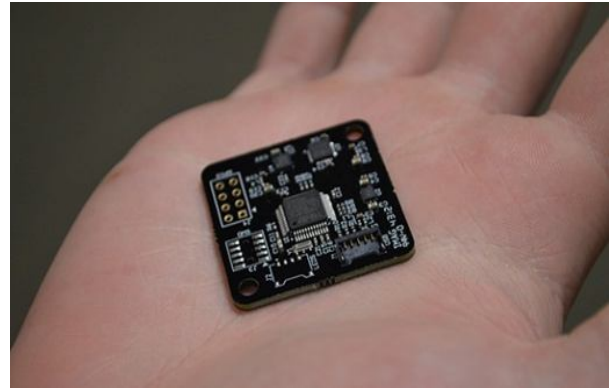

Fig.4 Head tracking sensor

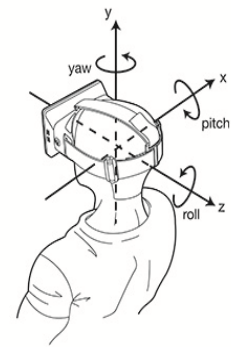

Fig5 An illustration of the three axes.

### B. Position tracking

Oculus rift DK2 involves position tracking which allows the user to lean in for a closer look at an in-game objectg or panel, or peek around a wall by moving their head and upper body, and seeing their physical actions translated into the virtual world. The added positional tracking helps with reducing the dizziness, the brain doesn't get confused by the missing degree of motion.

### C.   Control box

Oculus Rift comes with its own power supply of standard USB voltage of 5 V DC a. The power supply circuit is within the control box. The control box is the interface between headset and the computer. The basic control is done by control box. There are five control buttons on it indicating contrast, brightness and power. It has various ports such as Digital Video Interactive (DVI), High Definition Multi-media Interface (HDMI), mini-USB and DC power connection ports**.**

## V. INTERFACING

The Oculus SDK currently supports MacOS, Windows (Vista, 7, 8) and Linux. There are no specific computer hardware requirements for the Oculus SDK; however, a computer with a modern graphics card is recommended. A good benchmark is to try running Unreal Engine 3 and Unity at 60 frames per second(FPS)with vertical sync and stereo 3D enabled. The following components are provided as a guideline:

- MacOS: 10.6+
- Linux: Ubuntu 12.04 LTS
- 2.0+ GHz processor
- 2 GB system RAM
- Direct3D10 or OpenGL 3 compatible video card
- Windows: Vista, 7, or 8

Although many lower end and mobile video cards, such as the Intel HD 4000, have the graphics capabilities to run minimal Rift demos, their rendering throughout may be inadequate for full-scene 60 FPS. The System Development Kit also supports gaming controllers which include Xbox 360 wired controllers for Xbox and the Sony play station. DUAL shock3 controller for MacOS. To use the headset the control box is connected with the computer via the USB port using HDMI or DVI. The Oculus SDK is publicly available and it is an Open source

## VI. SCREEN DOOR EFFECT AND GHOSTING

Screen door effect can be described as a black grid over the original image. This occurs because of empty spaces in between the pixels. The display has a characteristic known as the Pixel Fill Factor that is responsible for this effect. On any LCD display every pixel is made of three sub pixels, namely red green and blue. Human perception of different colors on the display is a result of the varying intensities of these sub pixels. The distance between these sub pixels is called as the pixel pitch which ultimately decides the pixel fill factor. Higher the pixel pitch higher is the pixel fill factor. The oculus rift has a fair pixel fill factor however since it is worn close to the eyes it gives rise to the screen door effect.

Ghosting is the appearance of faded trails behind any moving object. In oculus rift, the slow pixel switching time causes ghosting i.e. the pixels take a fair amount of time to change intensities as compared to the motion of the head. Faster the movements greater the ghosting, since the pixel switching lags behind. Thus ghosting persists until the head stops moving. This causes blurring of the scene. Ghosting can be avoided with a higher switching rate for the pixels.

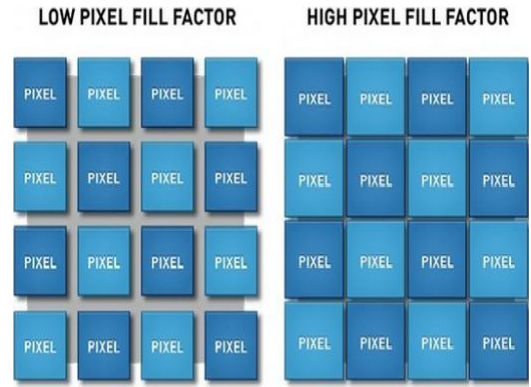

Fig.6 Pixel fill factor

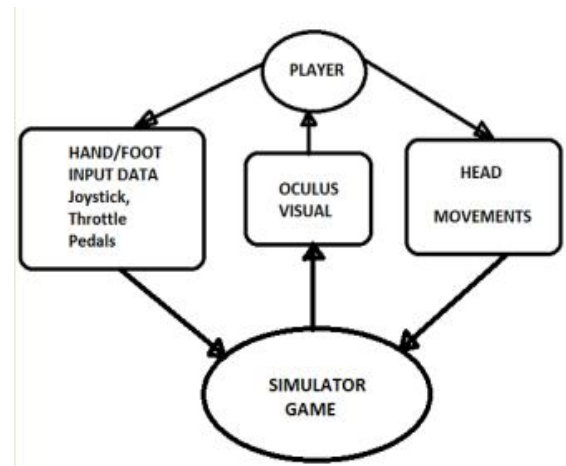

Fig.7 A block diagram illustrating the basic working of the oculus rift

## VII. RESULTS

Oculus Rift has an wide field of view, high resolution display. The Rift provides a truly immersive experience that allows you to step inside your favorite game.

## VIII. CONCLUSION

Since few years computer animation seems to be more based on dynamic simulation methods and physics. With the advent of VR, it feels like we are in 3-D computer generated environment. In future, real time complex animation system will be developed taking advantage using VR-devices like oculus rift and various simulation techniques. Oculus Rift is a virtual reality (VR) headset designed specifically for video games that will change the way you think about gaming forever.

## IX.     FUTURE SCOPE

Virtual reality is being used by pilots in the training process to help them overcome the fear of heights. With the help of VR is it also possible for scientists to comprehend the chemical reactions in a much better way and help them gather the minute details of any reaction. Architects can make an optimum use of VR by visualizing a building that is yet not into existence and to experience a virtual tour of the structure. Virtual reality also offers great deal of amusement with the museums and galleries. It is possible to view and experience a walk around in a museum with the help of oculus rift headsets and VR. For this purpose the museums should be recreated online which can then be accessed by the user for a real life experience. These could also be used by surgeons while in a surgery.

Virtual reality glasses such as the Oculus Rift will probably become cost efficient in the years to follow. Thus having them at the disposal of human benefit.

## X.     ACKNOWLEDGEMENT

We would like to thank the respected principal Dr. Hari Vasudevan of D. J. Sanghvi College of Engineering and Head of Department of Electronics and Telecommunication, Dr. Amit Deshmukh for giving us facilities and providing a propitious environment for working in the college. We would also like to thank S.V.K.M. for encouraging us in such co-curricular activities.

## REFRENCES